\documentstyle[amssym,epsfig]{mn}
\def\etal{et al.}
\def\planck{{\it Planck }}
\def\wmap{{\it WMAP }}

\title{ Late reionizations of the Universe and their manifestation in the
{\it WMAP} and future {\it Planck} data}

\author[Naselsky \& Chiang] {Pavel Naselsky$^{1-3}$ and Lung-Yih Chiang
$^{1}$\\ 
$^1$ Theoretical Astrophysics Center, Juliane Maries Vej 30,
DK-2100,  Copenhagen, Denmark \\ 
$^2$ Rostov State University, Zorge 5, 344090 Rostov-Don, Russia \\
$^3$ Niels Bohr Institute, Blegdamsvej 17, DK-2100 Copenhagen,
Denmark \\}

\date{Accepted 2003 ???? ???; Received 2003 ???? ???}

\begin{document}
\maketitle

\begin{abstract}
We investigate two sets of two-epoched reionization models and their
manifestation in the CMB anisotropy and polarization data of the recent
\wmap project and make some predictions for the future {\it Planck} missions. In the first set of models, the universe was
reionized twice, first at $z \simeq 15$ by population III stars and
then at  $z \simeq 6$ by stars in large galaxies. In the second set of
models, the extra peak-like reionization at high redshifts $z >100$ is
induced by the decay of unstable particles, followed by the standard
picture of reionization at $z\simeq 6$. We examine the general
properties of these two-epoched reionization models and their
implication in the {\it WMAP} CMB temperature anisotropy-polarization
cross-correlation. We have shown that these models have comparable
likelihood values for the {\it WMAP} data and distinct characters
which can be tested with the {\it Planck} high sensitivities.  

\end{abstract}


\newcommand{\nc}{\newcommand}
\newcommand{\beq}{\begin{equation}}
\newcommand{\eeq}{\end{equation}}
\newcommand{\be}{\begin{eqnarray}}
\newcommand{\ee}{\end{eqnarray}}
\newcommand{\Odm}{\Omega_{\rm dm}}
\newcommand{\Ob}{\Omega_{\rm b}}
\newcommand{\Om}{\Omega_{\rm m}}
\newcommand{\nb}{n_{\rm b}}
\newcommand{\num}{\nu_\mu}
\newcommand{\nue}{\nu_e}
\newcommand{\nut}{\nu_\tau}
\newcommand{\nus}{\nu_s}
\newcommand{\mnus}{M_s}
\newcommand{\taus}{\tau_{\nu_s}}
\newcommand{\nnt}{n_{\nu_\tau}}
\newcommand{\rnt}{\rho_{\nu_\tau}}
\newcommand{\mnt}{m_{\nu_\tau}}
\newcommand{\tnt}{\tau_{\nu_\tau}}
\newcommand{\bi}{\bibitem}
\newcommand{\rar}{\rightarrow}
\newcommand{\lar}{\leftarrow}
\newcommand{\lrar}{\leftrightarrow}
\newcommand{\dm}{\delta m^2}
\newcommand{\mpl}{m_{Pl}}
\newcommand{\mbh}{M_{BH}}
\newcommand{\nbh}{n_{BH}}
\def\zrec{z_{\rm rec}}
\def\zreio{z_{\rm reion}} 
%\makeatletter
\def\kms{\ifmmode{{\rm km}\,{\rm s}^{-1}}\else{km\,s$^-1$}\fi}
\def\mpc{{\rm Mpc}}

\newcommand{\eq}{{\rm eq}}
\newcommand{\tot}{{\rm tot}}
\newcommand{\M}{{\rm M}}
\newcommand{\coll}{{\rm coll}}
\newcommand{\ann}{{\rm ann}}

\section{Introduction}
A detailed study of the ionization history of the Universe is
fundamentally important for our understanding of the properties of the
structure and evolution of the Universe, particularly the large-scale
structure and galaxy formation. Although the epoch of galaxy
formation is often referred as the {\it dark age} due to the difficulty in
direct observations, it is, nevertheless, feasible to investigate in details
the ionization history of the Universe through the cosmic microwave background
(CMB) anisotropies and polarization. The recent CMB experiments,
such as the {\it BOOMERANG} \cite{boomerang}, {\it MAXIMA-1}
\cite{maxima}, {\it CBI} \cite{cbi}, {\it VSA}
\cite{vsa}, {\it DASI} \cite{dasi}, and its polarization   
data \cite{dasipol1,dasipol2} have shed light on probing the
dark age of the Universe. In particular, the newly released \wmap 
temperature-polarization data \cite{wmapdata1,wmapdata2} provide a new way for
understanding of the very early stages of galaxy and star formation. We expect
to have the future \planck polarization data  with unprecedented
accuracy. The polarization power spectrum from these two missions will
therefore provide us the information about the kinetics of hydrogen
recombination and allow us to determine the parameters of the last
scattering surface and the ionization history of the cosmic plasma at
very high redshifts $z\sim 10^3$.

In the framework of the modern theory of the primary CMB anisotropy
and polarization formation, the theory of hydrogen recombination are
assumed to be a `standard' one. The classical theory of hydrogen
recombination for the pure baryonic cosmological model was developed
by Peebles \shortcite{peebles}, Zel'dovich, Kurt and Sunyaev
\shortcite{zeldovich}, and was generalized for non-baryonic dark
matter by Zabotin and Naselsky \shortcite{zabotin}, Jones and Wyse
\shortcite{jones}, Seager, Sasselov and Scott \shortcite{recfast},
Peebles, Seager and Hu \shortcite{psh}. This standard model of
recombination has been modified in various ways. 

First of all, there are some variants from the
standard hydrogen recombination model, namely, the delay and acceleration
of recombination at the redshift $\zrec\simeq 10^3$ due to energy
injection from unstable massive particles \cite{dn} or due to the lumpy
structure of the baryonic fraction of the matter at small scales
\cite{nn}, in which the typical mass of the clouds is of the order
$10^5-10^6 M_{\odot}$ (see Doroshkevich \etal 2003 and the references
therein). Secondly, the most crucial part of the ionization history of
the Universe is related to the large-scale structure and galaxy 
formation and is called late reionization. The model of the late
reionization is not yet well-established and needs further
investigations. 

The conventional view of the ionization history\footnote{In this paper we refer to ``reionization epoch(s)''
as epoch(s) with $x_e$ above the residual fraction ($\sim10^{-3}$) of ionization
from the recombination epoch.} is that cosmological hydrogen
became neutral after recombination at  $\zrec\simeq 10^3$ and was
reionized at some redshift $\zreio$, 
\begin{eqnarray}
\zreio=13.6\left(\frac{\tau_r}{0.1}\right)^{2/3}\left(\frac{1-\langle
Y_p\rangle} {0.76}\right)^{-2/3}  \nonumber  \\
\times\left(\frac{\langle\Ob
h^2\rangle}{0.022}\right)^{-2/3}\left(\frac{\Odm h^2}{0.125}\right)^{1/3},
\label{eq1}
\end{eqnarray}
where $\tau_r$ is the Thomson optical depth, $\Ob$ is the present
baryonic density scaled to the critical density, $\Odm$ is the
dark matter density, $h=H_0/100 \kms \mpc$ is the Hubble constant, $\langle
Y_p\rangle$ is the helium mass fraction of matter. 
Recently Cen \shortcite{cen} has proposed the model of the late
reionization with two epochs. Firstly, hydrogen was reionized at redshift 
$\zreio^{(1)}\simeq 15$ by  Population III stars and secondly at
$\zreio^{(2)}\simeq 6$ by stars in large galaxies. On the other hand,
we also discuss another distinct feature of reionization model, which
is called {\it the peak-like reionization}. This
shoot-up in the ionization fraction at $z >100$ can be induced by
energy injection into the cosmic plasma. 

These two-epoched reionization models, which can be tested by the 
\wmap and future \planck data \cite{cen}, would be significant for the
interpretation of the polarization measurements. The
polarization of the CMB from the late reionization epoch (or epochs)
is sensitive to the width of the period $\Delta \zreio$, when the
ionization fraction $x_e$ increases from the residual ionization
($x_e\sim  10^{-3}$) up to $x_e \sim  0.1-1$ \cite{cmbfast}. They 
can provide unique information about the physical processes induced by
complicated ionization regimes. The aim of the paper is to discuss the
distinct characters of these models in the light of recent \wmap data 
and to predict the peculiarities
in the polarization power spectrum induced from
both the Cen model \shortcite{cen} of the late reionization and the
extra peak-like reionization, taking into account the
properties and the sensitivities of the upcoming polarization measurements. 

\section{Phenomenology of the two-epoched late reionization}
\begin{figure}
\centering
\epsfig{file=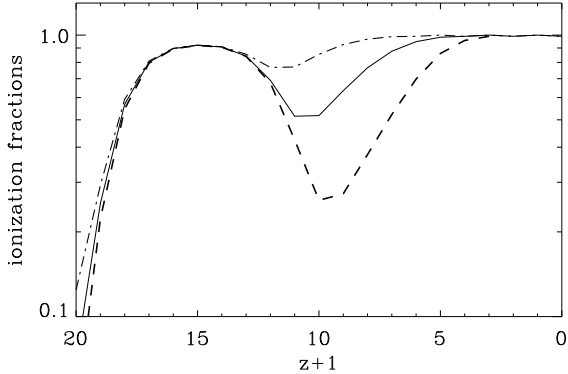,width=8.5cm}
\caption{
The ionization fraction for the
two-epoched reionization models. Solid line corresponds to the model 1,
the dash line the model 2 and the dash-dot line the model 3.
} \label{if}
\end{figure}

The model of the reionization process proposed by Cen \shortcite{cen}
can be described phenomenologically in terms of the injection of
additional Ly-$c$ photons via the approach by Peebles, Seager and Hu
\shortcite{psh}, Doroshkevich and Naselsky \shortcite{dn},
Doroshkevich \etal \shortcite{dnnn}. For the epochs of reionization
the rate of ionized photon production $n_{i}$ is defined as
\begin{equation}
\frac{dn_i}{dt}=\varepsilon_{i}(z) \nb(z) H(z),
\label{eq2}
\end{equation}
where $H(z)$ and $\nb(z)$ are the Hubble parameter
and the mean baryonic density at $z$, respectively,
$\varepsilon_{i}(z)$ is the effectiveness of the  Ly-$c$ photon production. 
%$t^{(1)}_{reio})$ corresponds to $z= z^{(1)}_{reio} $ and $\tau^2_1$ is the
%width of reionization.
As one can see from Eq.(\ref{eq:eq2}) the dependence of
$\varepsilon_{i}(z)$  parameter upon
redshift $z$ allows us to model any kind of ionization regimes, including
heavy particle decays. This parameter also includes uncertainties from the
fraction of baryons that collapse and form stars, and the escape fraction
for ionizing photons. For late reionization, the ionization
fraction of matter $x_e=n_e/\overline{n}$ can be obtained
from the balance between the recombination and the ionization process
\begin{equation}
\frac{dx_e}{dt}=-\alpha_{\rm rec}(T) \nb x^2_e +
\varepsilon_{i}(z) (1-x_e)H(z), 
\label{eq3}
\end{equation}
where $\alpha_{\rm rec}(T)\simeq 4\times 10^{-13} \left(T/10^4
K \right)^{-0.6} {\rm s}^{-1}{\rm cm}^{-3}$ is the recombination coefficient and $T$ is the
temperature of the plasma and $\nb$ is the mean
value of the baryonic number density of matter. In an equilibrium
between the recombination and the ionization process the ionization
fraction of the matter follows the well-known regime
\begin{equation}
\frac{x^{2}_e(z)}{1-x_e(z)} = \frac{\varepsilon_{i}(z) H(z)}{
\alpha_{\rm rec}(z) \nb(z)},
\label{eq4}
\end{equation}
where $H(z)=H_0\sqrt{\Om(1+z)^3+1-\Om}$ 
and $\nb \simeq 2 \times 10^{-7}(\Ob h^2/0.02)(1+z)^3$.
We would like to point out that Eq.(\ref{eq4}) can be used for any
models of the late reionization, including the Cen model
\shortcite{cen} by
choosing the corresponding dependence of the $\varepsilon_{i}(z)$
parameter on redshift. This point is vital in our modification of the
{\sc recfast} and the {\sc cmbfast} packages, from which we can use the
standard relation for matter temperature $T(z)\simeq 270
\left(1+z/100 \right)^2 {\rm K}$ and
all the temperature peculiarities of the reionization and clumping would be
related with the $\varepsilon_{i}(z)$ parameter through the mimic of
ionization history. For example, in the Cen model \shortcite{cen} the function
$T(t)$ has a point of maxima $T_{\rm max} \sim (1.3-1.5) \times 10^4$ at 
$z \sim \zreio^{(1)}$ and decreases slowly  at $z < \zreio^{(1)}$ down to
$T(t) \sim 10^4 \simeq {\rm const}$ at the redshift range
$6< z <12$. Let us introduce some model of the $\varepsilon_{i}(z)$
parameter dependence over $z$ as 
\begin{equation}
\varepsilon_{i}(z)= \varepsilon_{0} \exp
\left[-\frac{(z-\zreio^{(1)})^2}{\Delta z_1^2}\right]
+\varepsilon_{1}(1+z)^{-m}\Theta(\zreio^{(1)}-z), 
\label{eq5.0}
\end{equation}
where $\varepsilon_{0}$ ,$\varepsilon_{1}$ and $m$ are the free
parameters, $\Delta z_1\ll \zreio^{(1)}$ is the width of the first
epoch of reionization and $\Theta(x)$ is the step function. The first
term of Eq.(\ref{eq5.0})  corresponds
to peak-like reionization at $z=\zreio$, which decreases
significantly at $z > \zreio^{(1)}$. The second term is related to modelling the second
epoch of the reionization model discussed by Cen (2002), which results in
a monotonic increasing in $\varepsilon_{i}(z)$ function as a function
of time. From Eq.(\ref{eq5.0}) at $z\simeq \zreio^{(1)}$ we obtain
\begin{equation}
x_e\simeq 1-
\left(\frac{\varepsilon_{0} H(\zreio^{(1)})}{\alpha_{\rm rec}(\zreio^{(1)})
\nb(\zreio^{(1)})}\right)^{-1},
\label{eq5}
\end{equation}
where 
\begin{eqnarray}
\varepsilon_{0}& \gg & H^{-1}(\zreio^{(1)})
\alpha_{\rm rec}(\zreio^{(1)})\nb(\zreio^{(1)}); \nonumber  \\
&\simeq &
10^3\left(\frac{\Om h^2}{0.125}\right)^{-1/2}
\left(\frac{\Ob h^2}{0.022}\right)\left(\frac{1+ \zreio^{(1)}}
{16}\right)^{0.3}.  
\label{eq6}
\end{eqnarray}
One can find from Eq.(\ref{eq4}) and (\ref{eq5}) that for the second
epoch of reionization ($x_e\simeq 1$ at $z\simeq  \zreio^{(2)}\simeq 6$
in the Cen model \shortcite{cen}) the amplitude of the
$\varepsilon_{1}$ parameter and the index $m$ should satisfy the following
\begin{equation}
\varepsilon_{1}(1+\zreio^{(2)})^{-m} \sim \gamma \varepsilon_{0},
\label{eq7}
\end{equation}
where $\gamma$ is the parameter between 0.1 to 1 in order to 
model the properties of the Cen model and its variants.
We demonstrate the effectiveness of our phenomenological approach in
Fig.~\ref{if}: the ionization fraction $x_e$ against redshift for the three
models listed below:
\begin{itemize}
\item model 1: $\varepsilon_{0}=1.3 \times 10^3, \varepsilon_{1}=\beta
\times 10^9, \beta=1, m=7$;
\item model 2: $\varepsilon_{0}=1.3 \times 10^3, \varepsilon_{1}=\beta
\times 10^9, \beta=0.1, m=7$; 
\item model 3:  $\varepsilon_{0}=1.3 \times 10^3,
\varepsilon_{1}=\beta \times 10^9, \beta=100, m=8$. 
\end{itemize}
The curves are produced from the modification of the {\sc recfast} code
\cite{recfast}. Obviously the model 2 (dash line) is not physically viable, we
nevertheless can use it to test the sensitivity of CMB polarization to   
non-monotonic shape of ionization fraction. For all models we use the following values of the 
cosmological parameters: $\Ob h^2=0.022,\Om h^2=0.125, 
\Omega_{\lambda}=0.7, h=0.7, \Om + \Omega_{\lambda}=1$. We show the 3
models in Figure \ref{if}. Model 1 and 2 mimic the properties of
the Cen model (2002) where there is a dip in the reionization
fraction, whereas model 3 corresponds to roughly the standard reionization
model with no significant presence of peaks in ionization fraction.

\section{The CMB polarization for the two-epoched late reionization
models} 
In order to find out how sensitive the polarization
power spectrum is to the two-epoched reionization models, we consider
phenomenologically the different variants of hydrogen reionization
models by modifying the {\sc recfast} \cite{recfast} and {\sc cmbfast} code
\cite{cmbfast}. In Fig.~\ref{pol} we plot the polarization power
spectrum $C_p(\ell)$ for the model 1,  2, the standard single
reionization model at $\zreio \simeq 6$ and 13.6. The difference
between model 1 and 2 mainly lies in between the multipoles $2< \ell <
30$. 

\subsection{The anisotropy and polarization in comparison with the \wmap 
data}
\begin{figure}
\centering
\epsfig{file=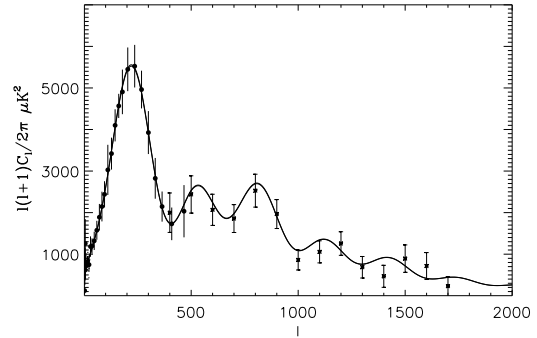,width=6.5cm}
\caption{
The anisotropy power spectrum for models $1-3$ in comparison with
\wmap and {\it CBI} data. All the models predict approximately the
same shape which cannot be seen in the figure due to degeneracy. The
corresponding likelihood parameters are shown in Table 1. The
\wmap data are taken from the official website at the
range $\ell \le 500$. The {\it CBI} data are taken  from {\it CBIM1}
and {\it M2} data sets. } \label{powerspectrum}
\end{figure}

\begin{figure}
\centering
\epsfig{file=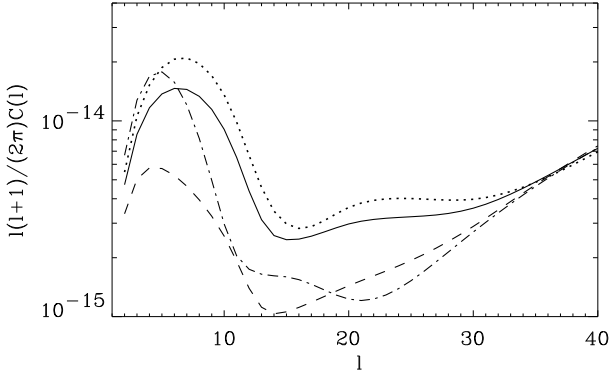,width=8.5cm}
\caption{
The polarization power spectrum for different models of 
the reionized universe. The solid line corresponds to the
model 1, the dotted line the model 2, the dash and the dash-dot line are
the model with single reionization ($x_e(\zreio)=1$) at $\zreio \simeq
6$, and at $\zreio \simeq 13.6$, respectively.} \label{pol}
\end{figure}

To characterize the differences between the reionization models we
have used the \wmap programme \cite{likelihood} for the
calculations of the likelihood for the   
anisotropy and $TE$ correlation power spectra against those from \wmap
results, shown in Table 1. As one can see, for all the models of late
reionization we get fairly good consistency for anisotropy and the $TE$
correlations. The accuracy of the \wmap data \cite{wmapdata1,wmapdata2}
however is not enough for discrimination between
the models. Therefore, the future \planck data is required for more
accurate investigations of the history of hydrogen reionization at
relatively low redshifts $z<30$. 
\begin{table} 
\centering 
\begin{tabular}{cccc} \hline 
Filter & & Cen model & \\ \hline 
model variants & 1 & 2 & 3 \\ \hline
Likelihood ($T$)& -494.357 & -525.928 & -495.159 \\ \hline 
Likelihood ($TE$)& -231.601 & -223.749 & -230.766 \\ \hline 
\end{tabular}
\caption{The likelihood parameters of the variants of the Cen model (2003). } 
\end{table}
For comparison, the \wmap best-fitting cosmological model has the 
likelihood parameter $-486.245$ (with 900 data points) for 
the anisotropy and $-228.695$ (with 499 data points) for the
$TE$ cross-correlation. As one can see from Table 1, all the models
have excellent agreement with the \wmap data, but we are not able to
distinguish with the accuracy of the \wmap data between the two-epoched
reionization models in the Cen model \shortcite{cen}. In order to
check the high multipole range of the power spectrum of temperature
anisotropies in the models $1-3$, in Fig.~\ref{powerspectrum} we plot
the $C(\ell)$, which are almost indistinguishable for all the models
and are consistent with the \wmap and the {\it CBI} data. The result shows
excellent agreements between the theoretical curves and the data points
from the experiments. 
The most intriguing question is could the future \planck data sets
allow us to distinguish any peculiarities of the late reionization epochs?

\subsection{The anisotropy and polarization in comparison with the future
\planck sensitivity}
The differences between the late reionization models in comparison
with the expected sensitivity of the \planck mission can be
expressed in terms of the power spectrum $C_p (\ell)$
(for the anisotropy, $E$ and $TE$ component of polarization) 

\begin{equation}
D_{i,j}(\ell)=\frac{2\left[C_{p,i}(\ell)-C_{p,j}(\ell)\right]}{C_{p,i}(\ell)+
C_{p,j}(\ell)}, 
\label{eqd}
\end{equation}
where the indices $i$ and $j$ denote the different models.

\begin{figure}
\centering
\epsfig{file=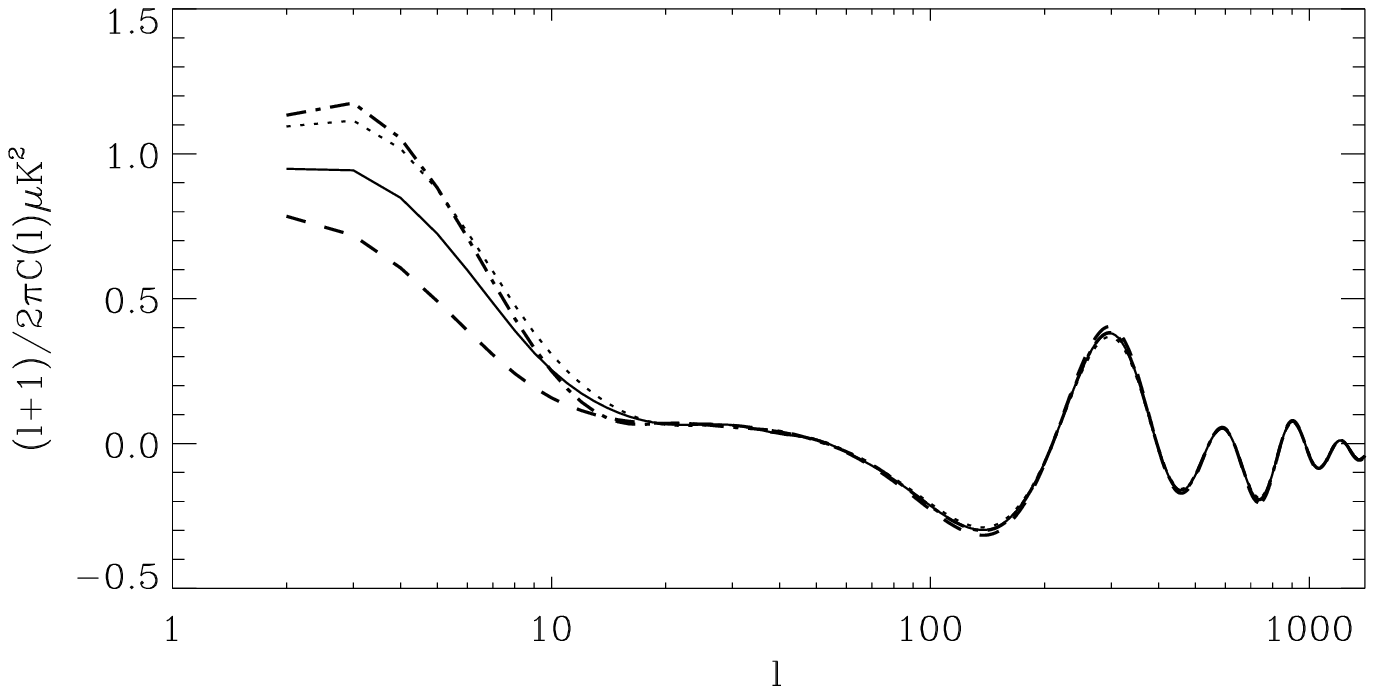,width=8.5cm}
\caption{
The $TE$ cross-correlation power spectrum for different models of 
the reionized universe. The solid line corresponds to the
model 1, the dotted line the model 2, the dash and the dash-dot line are
the model with single reionization ($x_e(\zreio)=1$) at $\zreio \simeq
6$, and at $\zreio \simeq 13.6$, respectively.} \label{te}
\end{figure}

\begin{figure}
\centering
\epsfig{file=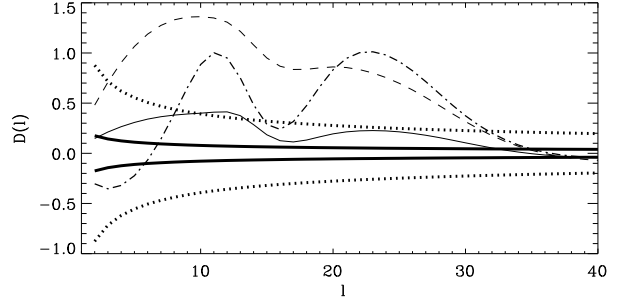,width=8.5cm}
\caption{The plot of $D_{i,j}$, the deviation in polarization for
different reionization models.
The solid line $D_{2,1}$ corresponds to the models 1 and 2, the dash-dot line
$D_{{\rm s},1}$ is from the model 1 in comparison with the model of single
reionization at $\zreio \simeq 13.6$ and the
dash line $D_{{\rm s},1}$ is from the model 1 in comparison with the 
model of single reionization at $\zreio \simeq 6$. The thick solid
(binning) and dotted curves ($f_{\rm sky}=0.65$, no binning) mark the
estimated  \planck errors.} \label{diff}
\end{figure}

In order to clarify the manifestations of the complex
ionization regimes in the models 1 and 2 we need to compare the peak to peak amplitudes of the
$D_{i,j}(\ell)$ function with the expected error of the anisotropy
power spectrum for the \planck experiment. We assume that
the systematics and foreground effects are successfully removed. The
corresponding error bar should be  
\begin{equation}
\frac{\Delta C(\ell)}{C(\ell)} \simeq \frac{1}{\sqrt{f_{\rm
sky}(\ell+\frac{1}{2})}}\left[1+ w^{-1}C^{-1}(\ell)W^{-2}_{\ell}\right],
\label{c}
\end{equation}
where $w=(\sigma_{p}\theta_{\rm FWHM})^{-2}$, $W_{\ell}\simeq
\exp\left[-\ell(\ell+1)/2\ell^2_s\right]$,
$f_{\rm sky} \simeq 0.65$ is the sky coverage during the first year of
observations, $\sigma_{p}$ is the sensitivity per resolution element
$\theta_{\rm FWHM} \times \theta_{\rm FWHM}$ and $\ell_s=\sqrt{8\ln 2}
\, \theta^{-1}_{\rm FWHM}$.

For all \planck frequency channels, for example, the FWHM
are less than 30 arcmin, so for the estimation of the errors at $\ell
\le 40$ range we can omit the second
term in Eq.~(\ref{c}). In Fig.3 we show the polarization
power spectrum and TE cross-correlation for models . While polarization
power spectrum is not observed by WMAP, the upcoming Planck mission will
be able to provide us the observation to differentiate different
models. It is clear from Fig.4 that the
TE cross-correlation spectra for different regimes of reionization have   
different shape at $\ell \le 10$ range but do not vary significantly
for $ \ell > 10$. All the deviations lie inside the cosmic variance
and are practically not observable.

As one can see from Fig.~\ref{diff} for $
D_{1,2}(\ell)$ the corresponding peak to peak amplitudes are on the order
of magnitude $20\%$ at $\ell \sim 10-40$, while the errors
$\Delta C(\ell)/C(\ell)$ are in about the same one. Such small deviations in
the polarization power spectrum caused by the
complicated ionization regimes can not be tested directly for each multipole
of the $C(l)$ power spectrum
by the \planck mission, even the systematic effects would
be removed down to the cosmic variance level.

As shown in Fig.~\ref{diff}, the deviation $D_{2,1}$ mostly lies
inside the error region. This indicates that the upcoming  \planck
observational data would not be able to distinguish the two-epoched
late reionization models from each other, where the only difference is
in the amplitudes of the minima of ionization fraction. However, it is
worth noting that both the models 1 and 2 have significant deviation
from the standard single reionization model (the dash and the dash-dot line). 

The shape of the polarization power spectrum in the two-epoched
reionization model differs from the shapes for single reionization
models even for $\zreio \simeq 13.6$. Such kind of dependence is
related to the difference with $\Delta z$  of the epoch when the
ionization fraction starts to grow from $x_{e, {\rm min}}\sim 10^{-3}$
up to  $x_{e, {\rm max}}\sim 1$.

An unique possibility to detect more complicated structure of late
reionization would be from the binning of the initial data, using,
for example, the same range of bin $\sim 15-30$. In such case, if
the correlations between each multipole are small, the accuracy of the
$C(\ell)$ estimation would be approximately $4-5$ higher then for unbinning
power spectrum and non-monotonic structure of the $D(\ell)$ function could be
detectable for the anisotropy and for $E$ and $TE$ polarization as well.

\section{Peak-like reionization at high redshifts}
In this section we shall investigate another type of two-epoched
reionization models. What is the implication of another reionization
occurring at high redshifts, if, for example, one of the epochs of the pre-reionization took place at redshifts $30 \ll z <1000$? Note that for
$z\gg 30$ the Compton cooling of the plasma is extremely important
and any energy injection to the cosmic plasma could produce relatively
short epochs of reionization, when the ionization fraction
became significantly higher ($x_{e, {\rm max}}\sim 1$), but for relatively
short time interval. We call such distinct character of reionization
{\it the peak-like reionization}. Such regimes can be
induced by the decay or annihilation of some unknown particles ore decay
of the primordial black holes \cite{naselsky,inn,kn} during the long period
$3 \times 10^5-10^8$ years. Because of the Compton cooling of the
plasma the injected energy density $\epsilon_{\rm inj}$ would be absorbed by
the CMB photons leading to $y$-distortion in the black-body spectra.
Peebles, Seager and Hu \shortcite{psh} have shown that the
corresponding value of the  $y$-parameter in this model is $y \sim
0.25 \epsilon_{\rm inj}/\epsilon_{\rm CMB}$, where $\epsilon_{\rm
CMB}$ is the energy density of the CMB at the redshift of the
injection. Taking into account the {\it COBE} upper limit for
$y$-parameter \cite{fixen} $y_{cobe}<2\times 10^{-5}$, one can
estimate the upper limit of the energy injection $\epsilon_{\rm inj}<
4 y_{cobe}\epsilon_{\rm CMB}$. On the other
hand, for reionization of each hydrogen atom we need to have roughly
one photon with energy $E \simeq I$, where $I \simeq 13.6 {\rm eV}$. Thus,
$\epsilon_{\rm inj}\sim x_e I \nb$ and we obtain the limit
\begin{eqnarray}
x_e \le 4y_{cobe}\left(\frac{\epsilon_{\rm CMB}}{I \nb}\right)=
4y_{cobe}\left(\frac{m_p}{I} \right) \left(\frac{\epsilon_{\rm CMB}}
{\epsilon_{\rm b}}\right) \nonumber  \\
\sim  10^2 (1+z_{\rm inj})\left(\frac{\Ob h^2}{0.022}
\right)^{-1} 
\label{c1}
\end{eqnarray}
where $m_p\simeq 1$ GeV is the proton mass and $\epsilon_{\rm b}$ is the
energy density of baryons at redshift $z_{\rm inj}$. From
Eq.(\ref{c1}) it is clear that the peak-like reionization ($x_e \sim
1$) is self-consistent with the {\it COBE} observational limit on
$y$-parameters. We can describe the peak-like reionization in terms
of the injection of an additional Ly-$c$ photons as in Section 1, but
with the source term now the Dirac $\delta$-function
\begin{equation}
\frac{dn_i}{dt}=\eta \nb(z)\delta_D(t-t_p),
\label{eqc2}
\end{equation}
where $\eta$ is the effectiveness of the Ly-$c$ photon production,
and $t_{p}$ is the age of the Universe at the moment of peak-like
reionization. Thus for the ionization fraction {\bf $x_e=$} we get
\begin{equation}
\frac{dx_e}{dt}=-\alpha_{\rm rec}(T)\nb x^2_e + \eta (1-x_e)\delta(t-t_p).
\label{eqc3}
\end{equation}

\begin{figure}
\centering
\epsfig{file=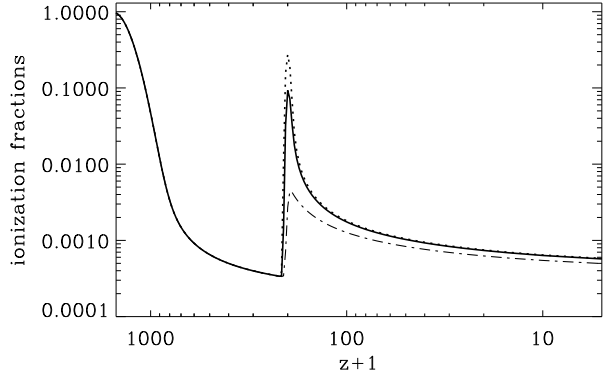,width=8.5cm}
\caption{
The ionization fraction for the peak-like reionization model (model
4). The dotted, the solid and the dash-dot lines correspond to model 4a,
4b and 4c.} \label{ifpeak}
\end{figure}

Quantitatively we can assume that for the first reionization epoch the
maximum of the ionization fraction is small ($x_e \ll 1$) and we neglect
$x_e$ in the $1-x_e$ term in Eq.(\ref{eqc3}). In such an approximation
the ionization balance of plasma is determined by the pure balance
between the recombination term and the energy injection term in
Eq.(\ref{eqc2}): 
\begin{equation}
x_e(t)\simeq \eta\left(1+\eta \int\limits_{t_p}^t \alpha(T) \nb
dt\right)^{-1},  
\label{eqc4}
\end{equation}
for the temperature of the plasma we can assume that $T(t_p)\sim (1-2)
\times 10^4$ K and at $t>t_p$, 
\begin{equation}
\frac{dT}{dt}=\frac{T_{\rm CMB}- T}{\tau_c},
\label{eqc5}
\end{equation}
where $\tau_c$ is the Compton cooling time. Note that our assumption
about maxima of $T(t_p) \sim 10^{-1} {\rm K}$ is closely related with
the energy balance. For reionization of the hydrogen by
electromagnetic cascades the typical energy release is in the order $I
\nb x_e$. Such a part of the energy of a cascade does not behave as
a simple $\delta$-function in the energy spectrum (see Doroshkevich \&
Naselsky 2002) and is usually characterised by power law
$E^{-\gamma}$. Because of Compton scattering of high energy
$\gamma$-quanta off electrons, it is natural to estimate the kinetic
energy of electrons as $\nb k T \sim I 	\nb x_e$ and $T \sim I x_e/k
\sim 10^{4} {\rm K}$ where $k$ is the Boltzmann constant. At $z > 40$ the ratio
between characteristic time of recombination $t_{\rm
rec}=[\alpha(T=10^4{\rm K})\, \nb]^{-1}$ and $\tau_c$ is more than one
order of magnitude, while both of them are
much smaller than the Hubble time. The relaxation of the
matter temperature to the CMB temperature proceeds faster than the
ionized hydrogen becoming neutral. Thus, while the temperature of
matter is close to the CMB temperature $T_{\rm CMB}$, the
corresponding time of recombination is  
\begin{equation}
\Delta t_{\rm rec}\simeq \frac{x_e}{|dx_e/dt|} \simeq
\eta^{-1} t_{r}(T_{\rm CMB}),
\label{eqc6}
\end{equation}
where $t_r=t_{\rm rec}$ at $T=T_{\rm CMB}$.
Taking into account the above-mentioned properties of the temperature
history of the plasma we can estimate the Thomson optical depth $\Delta
\tau_r$ caused by the peak-like reionization as follows
\begin{equation}
\Delta\tau_r\simeq  \tau_r \left(\frac{ t_{r}}{t_p}\right) \ln
\left(1+\eta\frac{t_p}{t_{r}}\right), 
\label{eqc7}
\end{equation}
where $\tau_r$ satisfies Eq.(\ref{eq1}) at $\zreio=\zreio(t_p)$.
For example, if the peak-like reionization takes place at the redshift
$\zreio=200$  and $\eta \simeq 0.1-0.2$ then 
\begin{equation}
\frac{t_r}{t_p}\simeq 0.1 (1+\zreio)^{-0.9}\left(\frac{\Ob h^2}{0.022}
\right)\left(\frac{\Om h^2}{0.125}\right)^{-1/2},
\label{eqc8 }
\end{equation}
and the corresponding values of the optical depth from Eq.(\ref{eqc7}) are
in order of the magnitude $\Delta\tau_r\simeq 0.03-0.05$. As one can
see, if the CMB anisotropy
data are consistent with the limit on Thomson optical depth of reionization
$\tau_r\le 0.1$ \cite{dnnn}, then roughly $30-50\%$ of  
$\tau_r$ can be induced by the peak-like reionization and $50-70\%$ with
the late reionization caused by the structure formation at low redshifts.
To describe the peak-like reionization numerically we use a Gaussian
approximation for the energy injection in Eq.(\ref{eqc2})
\begin{equation}
\eta \delta(t-t_p)\rightarrow \xi H(z)
\exp \left[-\frac{(z-\zreio)^2}{(\Delta z)^2}\right]
\label{eqc9}
\end{equation}
and describe the following two sets of examples for the peak-like reionization:
\begin{itemize}
\item model 4a: $\zreio=200$ with $\Delta z=5$, $\xi=100$; 
\item model 4b: $\zreio=200$ with $\Delta z=5$, $\xi=10$; 
\item model 4c: $\zreio=200$ with $\Delta z=5$, $\xi=1$,
\end{itemize} 
and 
\begin{itemize}
\item model 5a: $\zreio=500$ with $\Delta z=12.5$, $\xi=100$;
\item model 5b: $\zreio=500$ with $\Delta z=12.5$, $\xi=10$;
\item model 5c: $\zreio=500$ with $\Delta z=12.5$, $\xi=1$,
\end{itemize}
where $\Delta z/\zreio=0.025$ for both sets of models. The late
reionization $x_e(z \simeq 6)=1$ at $z \simeq 6$ is included in both
model 4 and model 5, as it is the standard part of the {\sc cmbfast}
package. In
Fig.~\ref{ifpeak} we plot the shape of the ionization fraction of the
plasma $x_e$ for the model 4. As an
analytical description, one can see the peaks of $x_e$ at
$\zreio=200$, which drops down at $z \sim 100-150$.

\begin{figure}
\centering
\epsfig{file=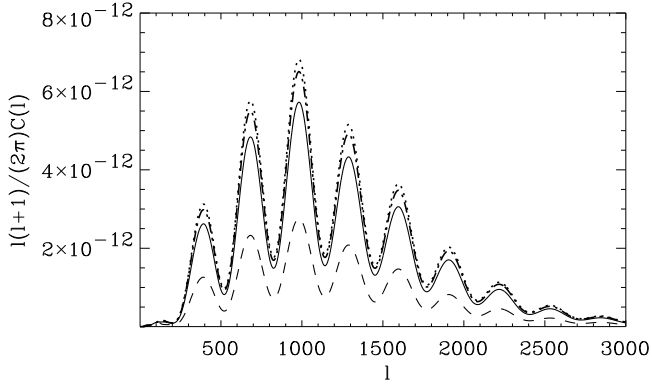,width=8.5cm}
\caption{The polarization power spectrum for the extra peak-like
reionization models (the model 4). 
The dash, the solid and the dash-dot line correspond to the model 4a, 
4b and 4c, respectively. The dotted line is the standard single reionization 
model at $\zreio \simeq 6$. 
} \label{pps} 
\end{figure}

\begin{figure}
\centering
\epsfig{file=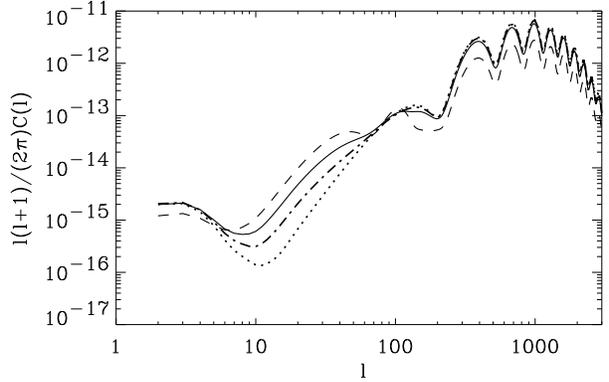,width=8.5cm}
\caption{The polarization power spectrum for the model 4 as in 
Fig.~\ref{pps} but in logarithmic scale, in which the differences 
between the power spectra can be seen more clearly at the multipole range 
$\ell <100$.
} \label{log1}
\end{figure}

\begin{figure}
\centering
\epsfig{file=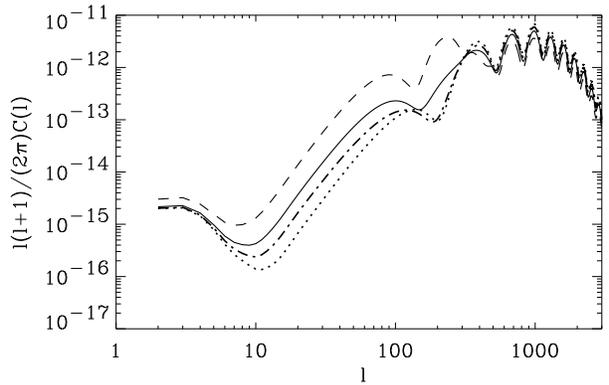,width=8.5cm}
\caption{The polarization power spectrum for the extra peak-like 
reionization models (the model 5) in logarithmic scale. The dash, 
the solid and the dash-dot line correspond to the model 5a, 5b and 5c, 
respectively. The dotted line is the standard single reionization model
 at $\zreio \simeq 6$.   
} \label{log2}
\end{figure}

\begin{figure}
\centering
\epsfig{file=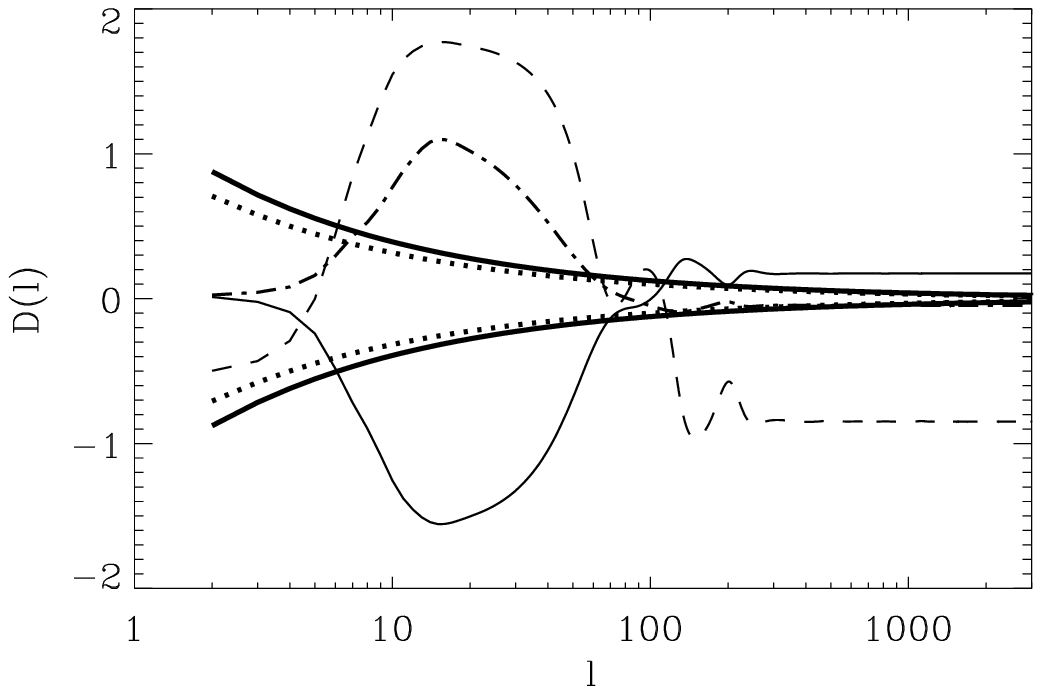, width=6.5cm}
\caption{
The plot of $D_{4,s}$ for comparison between the model 
4 and the
standard single reionization model at $\zreio \simeq 6$. The solid
line represents $D_{{\rm 4a},{\rm s}}$, the model 4a and the standard
single reionization model $\zreio \simeq 6$. The dash line 
$D_{{\rm4b},{\rm s}}$
 between the model 4b and the standard at $\zreio \simeq 6$
and the dash-dot line  $D_{{\rm 4c},{\rm s}}$ between 4c and the
standard at $\zreio \simeq 6$. The thick-dotted $(f_{\rm sky}=0.65)$ and 
solid lines $(f_{\rm sky}=1)$represent
the cosmic variance limit for  the \planck missions.  
} 
\label{pldiff1}
\end{figure}

\begin{figure}
\centering
\epsfig{file=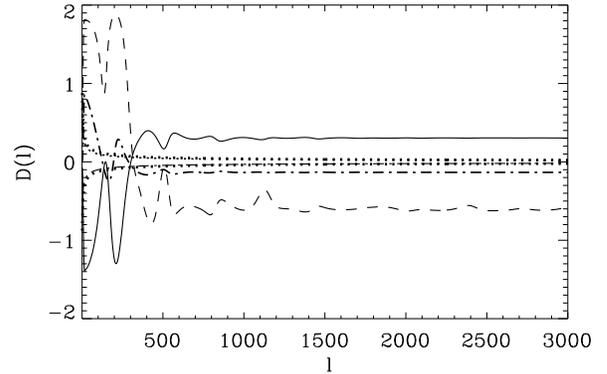,width=8.5cm}
\caption{
The plot of $D_{5,{\rm s}}$ for comparison between the model 5 and the
standard single reionization model at $\zreio \simeq 6$. The solid
line represents $D_{{\rm 5a},{\rm s}}$, the model 5a and the standard
single reionization model $\zreio \simeq 6$. The dash line $D_{{\rm
5b},{\rm s}}$ between the model 5b and the standard at $\zreio \simeq 6$
and the dash-dot line  $D_{{\rm 5c},{\rm s}}$ between 5c and the
standard at $\zreio \simeq 6$. The thick-dotted $(f_{\rm sky}=0.65)$ and 
solid lines $(f_{\rm sky}=1)$ represent
the cosmic variance limit for the \planck missions.  
} 
\label{pldiff2}
\end{figure}

By modification of the {\sc cmbfast} code for the primary polarization, we
plot in Fig.~\ref{pps}--\ref{log2} the corresponding power spectrum
for the model 4 and 5. As one can see from Fig.~\ref{pps} the common manifestation of the
extra peak-like and the standard late reionization produce interesting
features in the power spectrum. Namely, for high multipoles $\ell$ the
amplitude of the power spectrum decrease as $\exp(-\tau_r)$, while for the
multipole range $\ell < 100$ the manifestation of the peak-like
reionization is very clear, in particular at $10 < \ell <200$ from
Fig.~\ref{log1}. Figure~\ref{log2} shows how important
the peak-like reionization at  $\zreio=500$ could be for the
distortion at higher multipole range of the polarization power
spectrum. In Fig.~\ref{pldiff1} we plot the $D_{4,{\rm s}}$ function, the
comparison between the model 4 and the standard single reionization
model at $\zreio \simeq 6$. Fig.~\ref{pldiff2} is the comparison
between the model 5 and the standard single reionization
model at $\zreio \simeq 6$. In Table 2, we show the calculated likelihood for the anisotropy and $TE$ correlation power spectra from the model 5 against
those from \wmap results \cite{wmapdata1,wmapdata2}. They are close to the
parameters from the model 4.
\begin{table}
\centering
\begin{tabular}{cccc} 
\hline Filter & & peak-like model & \\ \hline 
model variants & 5a & 5b & 5c \\ \hline 
Likelihood ($T$)& ruled out & -770.693 & -509.583 \\ \hline 
Likelihood ($TE$)& -341.908 & -261.204 & -236.388 \\ \hline 
\end{tabular} 
\caption{The likelihood of the variants of the peak-like model (model 5).}
\end{table}

Once again we would like to point out that all the
peculiarities induced by the extra peak-like reionization have localized
structure which appears at some fixed multipole range. These features
can by tested by the \planck polarization measurements.

\section{Conclusions}
We have investigated the two-epoched reionization models of the
Universe. The two-epoched reionization can be induced by the structure
formation as described in Cen model \shortcite{cen}, or caused by unknown sources of
the energy injection (peak-like reionization) at relatively high
redshifts $z>30$. We have shown that for the Cen model \shortcite{cen}
the \wmap and
the \planck mission would be able to detect the general shape of the
ionization history for the two-epoched reionized plasma, which differs
from the single reionization models at $z \simeq 13.6$ or $z \simeq
6$. However, any peculiarities of the ionization fraction of the
matter inside the range $6 < z <13.6$, such as the decreasing of
ionization, do not  observed by the \wmap experiment due
 to the statistical significance
from the cosmic variance effect. The peak-like reionization model, on
the other hand, has some distinct features in the shape of
ionization fraction, and of the polarization power spectrum as
well. The most pronounced manifestation of the peak-like reionization
model is the localized features in the polarization power spectrum
which differs from the standard single reionization model. We reckon
that such kind of deviation from the standard reionization model,
in case of confirmation by  the \planck data, will be
significant for investigation of unstable particles or any relic
decaying during the `dark age' of the Universe.  

Note that in this paper we do not consider the secondary anisotropies and
polarization produced by the peak-like reionization at high
redshifts. These effects seem to be important if we take into account
the fact that the relaxation of the peculiar velocity of baryonic
matter and dark matter at  $z \simeq 200$ is completed and we can have
the analog of the Ostriker-Vishniak effect and the Doppler effect, but
for specific shape of the ionization fraction. These effects will be
investigated in the next paper.

\section*{Acknowledgments}
This paper is supported in part by
Danmarks Grundforskningsfond through its support for the
establishment of the Theoretical  Astrophysics Center.

\end{document}